\documentstyle[12pt,epsf]{article}
\newif\ifhypertex
\ifx\hyperdef\UnDeFiNeD
    \hypertexfalse
    \def\hyperdef#1#2#3#4{#4}
    
    \def\e@tf@ur#1{}
    \def\hth/#1#2#3#4#5#6#7{{\tt hep-th/#1#2#3#4#5#6#7}}
    \def\CERN{CERN, Geneva, Switzerland}
\else
    \hypertextrue
  \def\hth/#1#2#3#4#5#6#7{
  {\tt hep-th/#1#2#3#4#5#6#7}}
\def\CERN{
CERN, Geneva, Switzerland}
\fi
\makeatletter
\@addtoreset{equation}{section}
\makeatother
\renewcommand{\baselinestretch}{1.1}

\newcommand{\be}{\begin{equation}}
\newcommand{\ee}{\end{equation}}
\newcommand{\eel}[1]{\label{#1}\end{equation}}
\newcommand{\bea}{\begin{eqnarray}}

\newcommand{\eea}{\end{eqnarray}}
\newcommand{\eeal}[1]{\label{#1}\end{eqnarray}}
\newcommand{\beac}{\begin{equation}\begin{array}{rcl}}
\newcommand{\eeacn}[1]{\end{array}\label{#1}\end{equation}}

\renewcommand{\AA}{{\cal A}}

\newcommand{\CC}{{\cal C}}

\newcommand{\FF}{{\cal F}}

\newcommand{\MM}{{\cal M}}

\newcommand{\RR}{{\cal R}}
\newcommand{\SS}{{\cal S}}

\newcommand{\del}{\partial}

\newcommand{\non}{\nonumber}

\newcommand{\journal}[4]{{\em #1~}{\bf #2}\,(19#3)\,#4.}

\newcommand{\prl}{\journal {Phys. Rev. Lett.}}

\newcommand{\np}{\journal {Nucl. Phys.}}
\newcommand{\pl}{\journal {Phys. Lett.}}

\begin{document}
%
\begin{titlepage}
\topskip0.5cm
\hfill\hbox{CERN-TH/95-180}\\[-1.4cm]
\flushright{\hfill\hbox{hep-th/9507008}}\\[3.3cm]
\begin{center}{\Large\bf
On the Monodromies of $N\!=\!2$ Supersymmetric Yang-Mills Theory with
Gauge Group $SO(2n)$\\[2cm]}{
\Large A. Brandhuber and K. Landsteiner
\\[1.2cm]}
\end{center}
\centerline{\CERN}
\vskip2.cm
\begin{abstract}
We present families of algebraic curves describing the moduli space
of  $N\!=\!2$ supersymmetric Yang-Mills theory with gauge group
$SO(2n)$. We test our curves by computing the weak coupling
monodromies and the number of $N\!=\!1$ vacua.
\end{abstract}
\vfill
\hbox{CERN-TH/95-180}\hfill\\
\hbox{July 1995}\hfill\\
\end{titlepage}
%
%
\renewcommand{\baselinestretch}{1.2}\large\normalsize
\noindent{\bf Introduction}

A year ago Seiberg and Witten \cite{sw} showed how to obtain an exact
solution for the low energy effective action of $N\!=\!2$
supersymmetric Yang-Mills theory with gauge group $SU(2)$. Shortly
afterwards the generalization to gauge groups $SU(n)$ with arbitrary
$n$ has been obtained in \cite{klty,af}. For these cases a lot of
work has been done by now. The large $n$-limit and the connection to
$N\!=\!1$ theories has been studied in \cite{ds}. The
non-perturbative effective action has been elaborated in \cite{klt}.
Phases with mutually non-local massless dyons were identified in
\cite{ad}. Also $N\!=\!2$ supersymmetric QCD with matter in the
fundamental representation of the gauge group has been considered in
\cite{sqcd}. Fairly recently a solution for gauge groups $SO(2n+1)$
has been presented in $\cite{bu}$. The purpose of this letter is to
extend these previous results to gauge groups $SO(2n)$. It should be
noted here that in the Cartan classification the groups $SO(2n)$
correspond to simply laced Lie algebras of type $D_n$ whereas
$SO(2n+1)$ correspond to non-simply laced ones of type $B_n$.

\noindent{\bf Semiclassical Monodromies}

Following \cite{sw} we write the Lagrangian for $N\!=\!2$ SYM with
arbitrary gauge group $G$ in $N\!=\!1$ superspace language as
\be
{\cal L} = \frac{1}{4\pi} Im \left( \int d^4\theta
\frac{\del\FF}{\del\AA}
\bar\AA + \int d^2\theta \frac 1 2 \frac{\del^2 \FF}{\del\AA^2}
W_\alpha W^\alpha \right)
\eel{action}
where the $N\!=\!2$ vectormultiplet has been decomposed into an
$N\!=\!1$ chiral
superfield $\AA$ and an $N\!=\!1$ vectorfield $W_\alpha$. Two facts
are of utmost importance here. Firstly the whole theory is governed
by a single holomorphic function, the prepotential $\FF(\AA)$.
Secondly there is a nontrivial classical potential for the scalar
field, $V(\phi) = \left[\phi,\phi^\dagger\right]$. From this it
follows that for generic vacuum expectation value of $\phi$ the gauge
group is broken down to the maximal torus. Thus the low energy
degrees of freedom correspond to a $U(1)^r$-gauge theory with $r$
being the rank of $G$. Since $<\phi>$ can always be chosen to lie in
the Cartan sub-algebra, whose generators we denote by $H_i$, we have
$<\phi> = \sum_{i=1}^r \phi_i H_i$. The W-bosons corresponding to the
roots ${\vec{\alpha}}$ of the gauge group acquire a mass which is
proportional to $(\vec{\phi}.\vec{\alpha})^2$. Whenever the vacuum
expectation value of the scalar field is orthogonal to a root, a
W-boson becomes massless and therefore the low energy description is
no longer valid there. This information can be compactly encoded in
the zeroes of the Weyl group invariant ``classical'' discriminant:
\be
\Delta_{cl} = \prod_{\vec{\alpha}\in\Psi_+}
(\vec{\alpha}.\vec{\phi})^2.
\eel{disccl1}
Here $\Psi_+$ is the set of positive roots. Two vectors $\vec{\phi}$
should be identified if they differ by a Weyl transformation of the
gauge group. In the space of the $\phi_i$ the zeroes of the
discriminant coincide with the fixed points under the Weyl group
action.

Quantum mechanically the theory is characterized by a dynamically
generated scale $\Lambda$. It is weakly coupled if $<\phi>$ is large
in comparison to $\Lambda$, and therefore a perturbative description
is valid in this regime. The prepotential takes the form
\cite{klt,bu}
\be
\FF_{pert} = \frac i {4\pi} \sum_{\vec{\alpha}\in\Psi_+}
(\vec{\alpha}.\vec{\phi})^2
\log \frac{(\vec{\alpha}.\vec{\phi})^2}{\Lambda^2}.
\eel{prep}
It is clear that the logarithm gives rise to singularities if
$(\vec{\alpha}.\vec{\phi})=0$. $\FF_{pert}$ is not a single-valued
function of $\vec{\phi}$, and thus gives rise to non-trivial
monodromies when one encircles a singularity in the moduli space of
vacuum configurations. Actually what one is interested in are
monodromies acting on the vector $(\vec{\phi}^D,\vec{\phi})$ where
the dual variables are defined as $\phi^D_i = \del \FF / \del\phi_i$.
Thus for a gauge group of rank r the monodromy group will be a
subgroup of the group of duality transformations $Sp(2r,Z)$.

In the perturbative regime there will be $r$ simple monodromies
corresponding to the simple roots of the gauge group. They generate
all the other semiclassical monodromies by conjugation. The monodromy
matrix $M_i$ induced by the Weyl reflection on the root $\alpha_i$
from the prepotential $\FF_{pert}$ is
\be
M_i = \pmatrix{
1- 2 \frac{\vec{\alpha}_i\otimes\vec{\alpha}_i}{\vec{\alpha}_i^2}& -
\vec{\alpha}_i\otimes\vec{\alpha}_i \cr
0 & 1- 2 \frac{\vec{\alpha}_i\otimes\vec{\alpha}_i}{\vec{\alpha}_i^2}
\cr}.
\ee
To be specific we write down the simple monodromies for $SO(8)$. We
chose an orthogonal basis such that the four simple roots are given
by \cite{fu}
\beac
\vec{\alpha}_1 = (1,-1,0,0) ,&\vec{\alpha}_2 = (0,1,-1,0) ,\\
\vec{\alpha}_3 = (0,0,1,-1) ,&\vec{\alpha}_4 = (0,0,1,1) .
\eeacn{roots}
The four simple monodromies are
{\small
\beac
M_1 = \left (\matrix{
	0 & 1 & 0 & 0 & -1 & 1 & 0 & 0 \cr
	1 & 0 & 0 & 0 & 1 & -1 & 0 & 0 \cr
	0 & 0 & 1 & 0 & 0 & 0 & 0 & 0 \cr
	0 & 0 & 0 & 1 & 0 & 0 & 0 & 0 \cr
	0 & 0 & 0 & 0 & 0 & 1 & 0 & 0 \cr
	0 & 0 & 0 & 0 & 1 & 0 & 0 & 0 \cr
	0 & 0 & 0 & 0 & 0 & 0 & 1 & 0 \cr
	0 & 0 & 0 & 0 & 0 & 0 & 0 & 1 \cr
}\right ) &
M_2 = \left (\matrix{
	1 & 0 & 0 & 0 & 0 & 0 & 0 & 0 \cr
	0 & 0 & 1 & 0 & 0 & -1 & 1 & 0 \cr
	0 & 1 & 0 & 0 & 0 & 1 & -1 & 0 \cr
	0 & 0 & 0 & 1 & 0 & 0 & 0 & 0 \cr
	0 & 0 & 0 & 0 & 1 & 0 & 0 & 0 \cr
	0 & 0 & 0 & 0 & 0 & 0 & 1 & 0 \cr
	0 & 0 & 0 & 0 & 0 & 1 & 0 & 0 \cr
	0 & 0 & 0 & 0 & 0 & 0 & 0 & 1 \cr
}\right )\non
\eeacn{simplemono1}
\beac
M_3 = \left (\matrix{
	1 & 0 & 0 & 0 & 0 & 0 & 0 & 0 \cr
	0 & 1 & 0 & 0 & 0 & 0 & 0 & 0 \cr
	0 & 0 & 0 & 1 & 0 & 0 & 1 & -1 \cr
	0 & 0 & 1 & 0 & 0 & 0 & -1 & 1 \cr
	0 & 0 & 0 & 0 & 1 & 0 & 0 & 0 \cr
	0 & 0 & 0 & 0 & 0 & 1 & 0 & 0 \cr
	0 & 0 & 0 & 0 & 0 & 0 & 0 & 1 \cr
	0 & 0 & 0 & 0 & 0 & 0 & 1 & 0 \cr
}\right ) &
M_4 = \left (\matrix{
	1 & 0 & 0 & 0 & 0 & 0 & 0 & 0 \cr
	0 & 1 & 0 & 0 & 0 & 0 & 0 & 0 \cr
	0 & 0 & 0 & -1 & 0 & 0 & -1 & -1 \cr
	0 & 0 & -1 & 0 & 0 & 0 & -1 & -1 \cr
	0 & 0 & 0 & 0 & 1 & 0 & 0 & 0 \cr
	0 & 0 & 0 & 0 & 0 & 1 & 0 & 0 \cr
	0 & 0 & 0 & 0 & 0 & 0 & 0 & -1 \cr
	0 & 0 & 0 & 0 & 0 & 0 & -1 & 0 \cr
}\right )
\eeacn{simplemonos}
}
According to the Dynkin diagram of $D_4$, they fulfill the braid
group relations
\be
M_i.M_2.M_i = M_2.M_i.M_2
\ee
with $i=1,3,4$.

\noindent{\bf The Curves}

Our aim is to reproduce these monodromies from an algebraic curve.
There are several ways to arrive at the desired expression. First let
us calculate $P=det(x.1\!\!1-\phi)$ where the matrices are in the
fundamental representation of $SO(2n)$.
\be
P(x) = \prod_{i=1}^{n} (x^2-e_i^2)
\eel{poly1}
We distinguished here the roots $e_i$ of the polynomial $P$ from the
vacuum expectation values $\phi_i$. This is because in the quantum
case both can be identified only in the semiclassical regime. The
Weyl group of $SO(2n)$ is the semi-direct product of $\SS_n\!$
{\scriptsize $\rhd\hspace{-2mm}<$}$Z_2^{n-1}$.  The first factor are
permutations of $n$ elements and the second denotes simultaneous sign
changes of two elements $e_i$, $e_j$. This distinguishes the $D_n$
series from the non-simply laced $B_n$ case. In the latter one can
flip the signs of each $e_i$ individually. The important consequence
is that the $D_n$ series has an exceptional Casimir of order $n$
(cfg. (\ref{casimirs}) below). Clearly $P(x)$ is invariant under the
Weyl
group. Therefore we can equally well  write it in terms of gauge
invariant variables
\be
P(x) = x^{2n} - x^{2(n-1)} u_2 - x^{2(n-2)} u_4 - ... -
x^2 u_{2(n-1)} - t^2 \,,
\eel{poly2}
with
\bea
u_{2m} &=& (-1)^{m-1} \sum_{i_1<i_1<..<i_m} e_{i_1}^2 e_{i_2}^2 ..
e_{i_m}^2\\
t &=& (i)^{n-1}\prod_{i=1}^{n} e_i\,.
\eeal{casimirs}
Note that due to the structure of the Weyl group the polynomial $P$
depends quadratically on $t$!

It has been suggested in \cite{klt} that the algebraic curves
describing the moduli space of $N\!=\!2$ SYM with gauge groups based
on the simply laced Lie algebras of ADE-type should be intimately
related to the simple singularities of the same type. Indeed the
$SU(n)$ curves are determined by the ``LG-potentials'' of type
$A_{n}$. Also the curves given in \cite{bu} are determined in exactly
the same manner by the well known potentials for the boundary
singularities $B_n$ \cite{arn}.
How can we establish such a relationship for the case at hand?

The perturbed simple singularities for the $D_n$ series is given by
\be
W_{D_n} = x_2^2 x_1 + x^{n-1} - x_1^{n-2} u_2 - x_1^{n-3} u_4 - ... -
u_{2(n-1)} - x_2 t\,.
\eel{lg}
Formally integrating out the variable $x_2$ by its equation of motion
and then substituting $x_1 \rightarrow \hat{x}^2$ one arrives at
\be
\bar{W}_{D_n} = \hat{x}^{2(n-1)} - \hat{x}^{2(n-2)} u_2 - ... -
u_{2(n-1)} - \frac{t^2}{\hat{x}^2}\,.
\eel{onex}
This is in fact well-known in the context of conformal field theory
\cite{dvv}.
Then our polynomial is simply $\hat{x}^2 \bar{W}_{D_n}$. At this
stage it is useful to have an explicit look at the discriminant of
$P$. To be specific, we chose the simplest generic case $D_4$ and
change notation according to $u_2 \rightarrow u$, $u_4 \rightarrow
v$, $u_6 \rightarrow w$
\beac
\Delta_{cl} &=& t^2 \Delta_W\\
\Delta_W &=& (256 t^6 + 27t^4u^4 + 144t^4u^2v + 128t^4v^2 +
4t^2u^2v^3 + 			16t^2v^4 + \\
& &\; +192t^4uw - 18t^2u^3vw - 80t^2v^2w - 6t^2u^2w^2 - 144t^2vw^2 -
		u^2v^2w^2 - \\
& &\;-4v^3w^2 + 4u^3w^3 + 18uvw^3 + 27w^4)\,.
\eeacn{disccl}
Here $\Delta_W$ is the discriminant that one computes from the
Landau-Ginzburg potential (\ref{lg}) or from the one-variable
expression (\ref{onex}). Using the basis of roots (\ref{roots}) one
finds that it precisely reproduces (\ref{disccl1}). Thus our
polynomial $P$ has an additional and a priori unexpected singularity
at $t=0$. We will see in the following how this puzzle resolves.

Following the standard arguments given in \cite{klty,af}, we suggest
for the quantum theory the following family of algebraic curves
\be
y^2 = P(x)^2 - \Lambda^{(4n-4)} x^4\,.
\eel{curve}
The classical theory has a $U(1)_\RR$ symmetry under which the field
$\phi$ has charge $2$. In the quantum theory it is anomalous and
broken down to a discrete $Z_{4 C_\nu}$-symmetry \cite{sw}. Here
$C_{\nu}$ is the dual Coxeter number of the gauge group. In the case
of $SO(2n)$ it is given by $C_{\nu}=2n-2$. If we assign charge $2$ to
$x$ and $4n$ to $y$, we see that our curve has precisely the required
$Z_{4(2n-2)}$-symmetry. The right hand side of (\ref{curve})
factorizes into
\be
\CC_+.\CC_- = (P(x) + \Lambda^{2n-2}x^2).(P(x) -
\Lambda^{2n-2}x^2)\,.
\ee
{}From this we infer that similarly as for $SU(n)$, the quantum
discriminant factorizes as well. We see that $\CC_\pm$  has the form
of the classical polynomial if we substitute $u_{2n} \rightarrow
u_{2n}\pm \Lambda^{2n-2}$. The factors $\CC_\pm$ have common roots
only at a point in the moduli space where $x=0$ is a root. However
this case is already contained in $\Delta_\pm$ itself and therefore
no new singularity arises.
Thus the quantum discriminant is given by
\be
\Delta_{qu} = t^4.\Delta_{+}.\Delta_{-}\,.
\eel{discqu}
\noindent {\bf Strong and weak coupling monodromies}

An important feature of our curve is that there appear only even
powers of $x$. Therefore it is invariant under a $Z_2$-symmetry
acting on the complex $x$-plane $\Pi: x\rightarrow -x$. This implies
that the homology can be uniquely decomposed into two subspaces
$H^+\oplus H^-$ where $H^+$ is the space of cycles which are
invariant under $\Pi$ and $H^-$ is the anti-invariant subspace
\cite{arn}. Our choice of cycles is depicted in Fig.1.\\
\begin{figure}[h]
\hbox to\hsize{\hss
\epsfysize=7.8cm
\epsffile{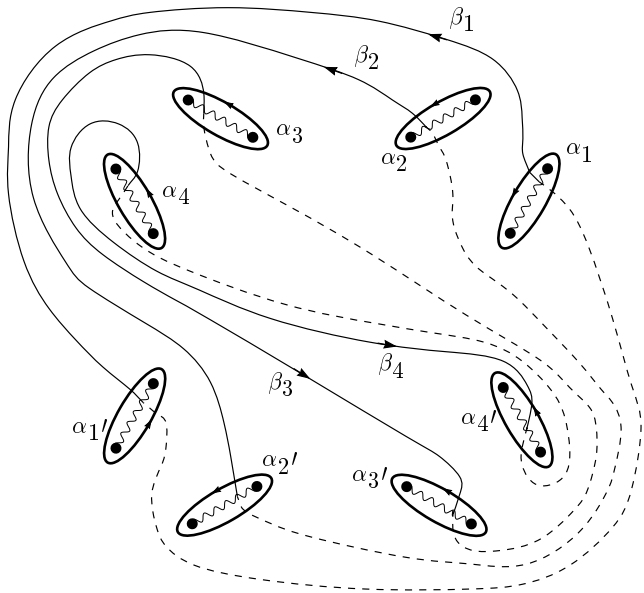}\hss}
\caption{The basic cycles for the $SO(8)$ curve}
\label{pic1}
\end{figure}

One sees that the cycles $\beta_i$ belong to $H^-$ whereas $\Pi
\alpha_i = \alpha_i\prime$. From the latter we define anti-invariant
cycles by taking the differences
\be
\Delta_i = \alpha_i - \alpha_i^\prime\,.
\ee
Cycles of this form are called long, cycles of the form of the
$\beta_i$ are called short. Their intersection form is given by
\be
<\Delta_i,\beta_j> = 2 \delta_{ij}\,.
\ee
In order to define the variables $e_i$ and their duals, $e^D_i$, as
period integrals, we also need a suitable meromorphic one-form
$\lambda$. For curves of the form (\ref{curve}) it has already been
derived in \cite{bu}. In our case it is given by
\be
\lambda = (2 P(x) - x P'(x))\, \frac {dx}{y}\,.
\eel{lambda}
Since it also changes sign under the action of $\Pi$, we obtain
invariant periods only by integrating over anti-invariant cycles.
Thus the physically relevant subspace of the homology is $H^-$. Now
the fields can be defined as
\bea
e_i = \oint_{\Delta_i} \lambda\,,&&e^D_i = \oint_{\beta_i} \lambda\,.
\eeal{periods}
An additional subtlety arises in considering the Picard-Lefschetz
formula. Since the intersection of any long anti-invariant cycle with
another anti-invariant cycle is always even we have to correct this
by a factor $\frac 1 2$. We obtain the modified Picard-Lefschetz
formula \cite{arn}
\bea
\delta\nu &=& \nu + \frac 1 2 <\nu,\mu_l>\mu_l\\
\delta\nu &=& \nu + <\nu,\mu_s>\mu_s\;,
\eeal{pl}
where $\mu_l$ is a long anti-invariant vanishing cycle and $\mu_s$ a
short one. It should be noted here that in the case of $SO(2n)$ we
will never encounter short anti-invariant vanishing cycles. This is
no surprise since these cycles correspond in a one-to-one manner to
short roots of Lie-algebras. Also we could have chosen a basis of
cycles consisting only of long anti-invariant ones. We did not do so
because the choice in Fig.1 is best suited to the orthogonal basis of
the Lie algebra. This will become clear from the discussion of the
strong coupling monodromies. In addition, it allows most easily to
compare
and to work out the differences to the case when the gauge group is
$SO(2n+1)$.

In order to derive the semi-classical monodromies
(\ref{simplemono1},\ref{simplemonos}), we use the fact that they can
be written as a product of two strong coupling monodromies. More
precisely, we have chosen a slice through the moduli space by fixing
$v$, $u$, $t$ and varying only $u$. Away from intersections or cusps
of the singular lines one sees four pairs of singular lines. Fixing a
base point in this plane we traced the effect on the branch points
when looping around the singularities. Depending on the particular
form of the loops, one finds different vanishing cycles. It suffices
however to choose eight loops, each encircling different singular
lines.\\
\begin{figure}[h]
\hbox to\hsize{\hss
\epsfysize=8cm
\epsffile{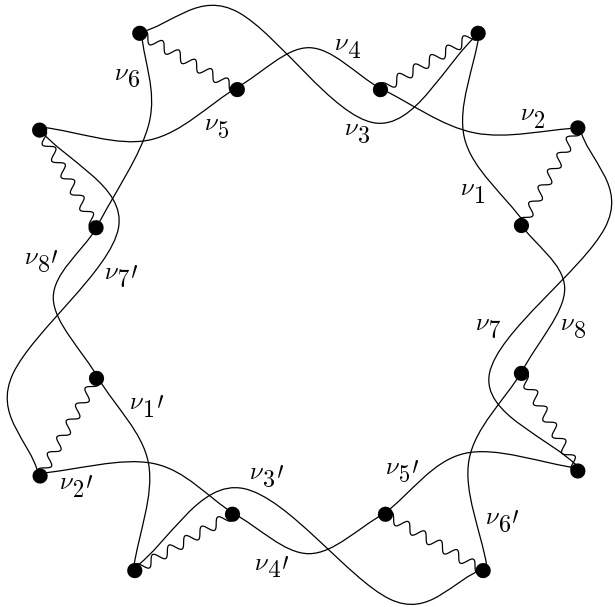}\hss}
\caption{The strong coupling vanishing cycles}
\label{pic2}
\end{figure}

All the other monodromies can then be obtained by conjugation from
the eight basic ones. The eight vanishing cycles can be seen in Fig.
2. Of physical relevance are again the differences $\delta_i =
\nu_i-\nu^\prime_i$. They can be expanded in terms of $\beta_i$ and
$\Delta_i$. We find
\beac
\delta_1 = (1,-1,0,0;0,0,0,0)\,,&\delta_2 =
(1,-1,0,0;1,-1,0,0)\,,\non\\
\delta_3 = (0,1,-1,0;0,0,0,0)\,,&\delta_4 = (0,1,-1,0;0,1,-1,0)\,,
\non\\
\delta_5 = (0,0,1,-1;0,0,-1,1)\,,&\delta_6 = (0,0,1,-1;0,0,0,0)\,,
\non\\
\delta_7 = (-1,0,0,-1;1,1,1,3)\,,&\delta_8 = (-1,0,0,-1;0,1,1,2)\,.
\eeacn{vancycl}
The first entries refer to magnetic quantum numbers and the others to
electric ones. A general formula for the monodromy matrix for a
massless dyon of magnetic charge $\vec{g}$ and electric charge
$\vec{q}$ has been given in \cite{klty}:
\be
\MM_{(\vec{g},\vec{q})}= \pmatrix{
1-\vec{g}\otimes\vec{q} & -\vec{q}\otimes\vec{q} \cr
\vec{g}\otimes\vec{g} & 1+\vec{g}\otimes\vec{q} \cr
}
\ee
Note that the magnetic quantum numbers of $\delta_{1,2}$ are given by
the root $\vec{\alpha}_1$, that of $\delta_{3,4}$ and that of
$\delta_{5,6}$ by
$\vec{\alpha}_2$ and $\vec{\alpha}_3$ respectively and finally that
of $\delta_{7,8}$ by the root
$-\vec{\alpha}_1-\vec{\alpha}_2-\vec{\alpha}_4$.
According to this we reproduce the weak coupling monodromies in the
following manner:
\beac
M_1 &=& \MM_{\delta_1}.\MM_{\delta_2}\non\\
M_2 &=& \MM_{\delta_3}.\MM_{\delta_4}\non\\
M_3 &=& \MM_{\delta_5}.\MM_{\delta_6}\non\\
M   &=& \MM_{\delta_7}.\MM_{\delta_8}\non\\
M_4 &=& M_1.M_2.M.M_2^{-1}.M_1^{-1}
\eeacn{monos}

Another nontrivial check of the curve (\ref{curve}) concerns the
quantum shift matrix $T$. It is obtained in the following way.
Performing the rotation $\Lambda^2 \rightarrow e^{2\pi i t}
\Lambda^2$, $t\in(0,1)$ one computes from  the prepotential
(\ref{prep}) the monodromy
\be
T= \left (\matrix{
1 & -\sum_{\vec{\alpha}\in\Psi_+} \vec{\alpha}\otimes\vec{\alpha} \cr
0 & 1  \cr
}\right )\,.
\ee
In the orthogonal basis we have $\sum_{\vec{\alpha}\in\Psi_+}
\vec{\alpha}\otimes\vec{\alpha} = C_\nu . 1\!\!1$. On the complex
curve this rotation induces pure braidings of the branch cuts. A
single braid of the cut $i$ induces $\beta_i \rightarrow \beta_i -
\Delta_i$. From the fact that $\Lambda$ appears to the power of two
times the dual Coxeter number we see that the number of braiding for
each branch cut is  $C_\nu$. Thus we find the quantum shift matrix in
the orthogonal basis.

\noindent {\bf The Monodromy around $t=0$ ?}

Now we come to a subtle and crucial point in our discussion. From the
expression of the quantum discriminant (\ref{discqu}) one expects
that some BPS state becomes massless as we approach $t=0$ for generic
values of $u$ , $v$ and $w$. A look at the algebraic curve
(\ref{curve}) tells us that four branch points i.e. two branch cuts
will collide at $x=0$ in this case, which seems to be a nasty,
unstable situation.
\begin{figure}[h]
\hbox to\hsize{\hss
\epsfysize=2.5cm
\epsffile{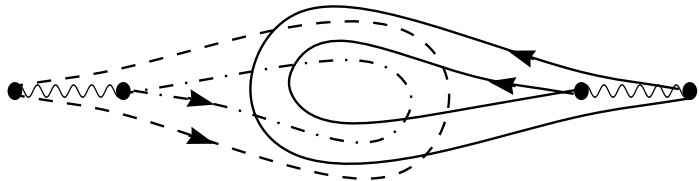}\hss}
\caption{Monodromy around $t = 0$ ?}
\label{pic3}
\end{figure}
But what does really happen as we loop around $t=0$~? We have checked
this situation for several one-dimensional slices through the moduli
space and in Fig. \ref{pic3} we have depicted what happens to the
branchpoints. Six of the branch cuts stay where they are, whereas two
such cuts rotate around each other and end up at their original
positions. Since no additional braidings occur, we see that the basic
cycles are unchanged and the associated monodromy matrix is simply
the identity. There are no massless particles associated with the
``singularity around $t=0$~'' !

This is in strong contrast to what is happening for the non-simply
laced gauge groups $SO(2n+1)$. In that case the monodromy around
$u_{2n} = 0$ \footnote{By $u_{2n}$ we mean the highest order gauge
invariant operator of $SO(2n+1)$, which is of order $2n$} has a
massless particle associated with it and one finds a strong coupling
monodromy. To be more specific we want to discuss $SO(5)$ in some
detail. We use basic cycles analogous to Fig. \ref{pic1} with
$\beta_2$ being $\zeta$ (they are the same as in \cite{bu}). The
algebraic curve and the quantum discriminant read as follows:
\be
y^2 = (x^4 - u x^2 - v)^2 - x^2 \Lambda^6 ~,
\ee
\be
\Delta_{qu}(u,v) = v^2(256 v^3 -128 u^2 v^2 + 16 u^4 v +
4 \Lambda^6 u^3 -144 \Lambda^6 u v -27 \Lambda^{12}) ~.
\ee

\begin{figure}[h]
\hbox to\hsize{\hss
\epsfysize=2.5cm
\epsffile{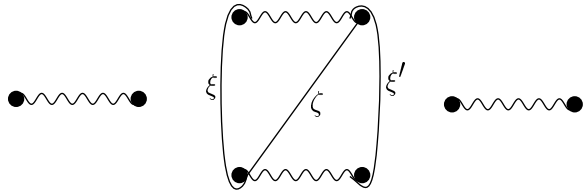}\hss}
\caption{ SO(5) vanishing cycles corresponding to a short root}
\label{pic5}
\end{figure}

In Fig. \ref{pic5} we have drawn two anti-invariant vanishing
cycles $\xi - \xi^\prime$ and $\zeta$ associated with a short root.

For the vanishing cycle $\xi-\xi^\prime$ we can calculate the
monodromy in the usual way and the magnetic-electric
quantum numbers read $\delta_{\xi-\xi^\prime} = (0,2;0,1)$.
Some difficulty arises for $\zeta$, which corresponds to a loop
around \footnote{$v$ is the highest order Casimir for SO(5)} $v=0$ in
the moduli space. In this case the two branch points encircled
by the vanishing cycle are not simply exchanged, but they rotate
around $x=0$ and move back to their starting points, picking up a
non-trivial braid. The monodromy ${\cal M}_\zeta$ obtained by using
the Picard-Lefschetz formula (\ref{pl}) is not the correct one,
because it corresponds to the exchange of the two branch points. In
fact one has to take its square ${\cal M}_\zeta^2$ with quantum
numbers $(0,2;0,0)$ i.e. the effective vanishing cycle is therefore
$2 \beta_2$. As expected, one obtains a weak coupling monodromy by
multiplying two strong coupling monodromies. ${\cal M}_\zeta^2 {\cal
M}_\xi$ is precisely the monodromy matrix $M_2$ of \cite{bu} and
corresponds to a Weyl reflection on a short root ! In the case of
$SO(2n+1)$ the singularity around $u_{2n} = 0$ is essential to
account for the presence of short roots. Of course, short roots are
not present for $SO(2n)$.

\noindent {\bf The $N\!=\!1$ Vacua}

Now we want to take a closer look at the moduli space of $SO(2n)$
gauge theories and the singularities living in it. The vanishing of
the quantum discriminant (\ref{discqu})\footnote{in the following
we will suppress the factor $t^4$ in (\ref{discqu})}
defines an $n-1$ dimensional singular submanifold in our $n$
dimensional moduli space. Since the quantum discriminant factorizes
into two pieces, one expects to find at least two branches of the
singular submanifold that intersect in some $n-2$-dimensional
manifold. Furthermore, the branches split themselves into several
branches and have self intersections and singularities, like cusps.
For arbitrary $n$ one will find a discrete set of points where $n$
branches intersect simultaneously. In the following we will
concentrate on these singular points, but it should be
mentioned that there are also higher dimensional singular
submanifolds which would be well worth to study.

At these singular points we can choose local coordinates and
the quantum discriminant will factorize into $n$ factors
(to lowest order). If some
of the factors are linear dependent, we are at a cusp singularity
where several of the $n$ massless dyons are mutually
non-local. Such points have been studied for the case of $SU(3)$ in
\cite{ds}. At some special points all branches intersect transversely
i.e. the factors are all linearly independent. All $n$ massless
particles are mutually local, and one can break $N\!=\!2$
supersymmetry down to $N\!=\!1$ by adding a mass term.

We have studied the case of $SO(8)$ in some detail, and found 48
points
where four branches of the singular manifold intersect. Mutually
non-local dyons are present at 42 of these points, whereas the
remaining
six points are the $N\!=\!1$ vacua
\be
u = 3~^3\sqrt{4} \Lambda^2 e^{2\pi \imath l/6}~,~
v = -\frac{9}{ ^3\sqrt{4}} \Lambda^4 e^{4\pi \imath l/6}~,~
w = \Lambda^6 e^{\pi \imath l}~,~t = 0
\eel{vacua}
with l = 0 \ldots 5. This is the correct number of $N\!=\!1$ vacua
for the $SO(8)$ gauge theory. We also checked this for $SO(6)$ and
found $4$ $N\!=\!1$ vacua.

For larger $n$ it becomes increasingly difficult to find the vacua
directly from the quantum discriminant. We propose therefore a
different approach which is similar to the discussion in \cite{ds}.
If we insert the values (\ref{vacua}) into our curve
(\ref{curve}) we find a polynomial with two simple roots, two double
roots and one sixfold root at $x=0$. This situation can easily
be generalized for arbitrary $n$ with the help of Chebyshev
polynomials.
The classical polynomial (\ref{poly2}) takes the form
\be
\hat{P_n}(x) = x^2 T_{2n-2}( \frac{x}{\Lambda} 2^\frac{1-2n}{2n-2} )
\Lambda^{2n-2}
\eel{cheby}
whereas the curve (\ref{curve}) will be of the form
\be
\hat{\CC_n}(x) = \hat{P}(x)^2 - x^4 \Lambda^{2n-2}~.
\eel{curve1}
We obtain a total of $2n-2$ solutions by complex rotation $x
\rightarrow x e^{\imath \pi l/(2n-2)}$ which is the correct number of
$N\!=\!1$ ground states. As a check one can compare the expansion
of (\ref{cheby}) for $n=4$ with the classical curve (\ref{poly2}),
and read off the values of $u$, $v$, $w$ and $t$. They perfectly
agree with the solutions found directly from the examination of the
singular points of the quantum discriminant of $SO(8)$ given in
(\ref{discqu}). We want to point out that the vacuum points found for
the $D_n$ series, with $t = 0$, coincide with the vacua for the
$A_{2n-3}$, series with all odd Casimirs set to zero.

\begin{figure}[h]
\hbox to\hsize{\hss
\epsfysize=4cm
\epsffile{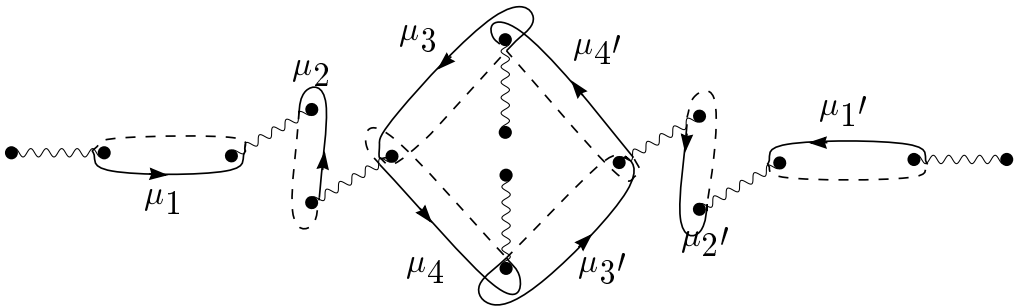}\hss}
\caption{Near a $N\!=\!1$ vacuum point one can see
two copies of the $D_4$ Dynkin diagram !}
\label{pic4}
\end{figure}

In Fig. \ref{pic4} we have depicted how the branch cuts arrange and
what cycles vanish as we approach an $N\!=\!1$ vacuum point. The
vanishing cycles have the usual anti-invariant form $\Delta\mu_i =
\mu_i - \mu_i\prime$. Apparently $\Delta\mu_1$ and $\Delta\mu_2$ are
non intersecting. For $\mu_3$ and $\mu_4$ the situation
is not so obvious but the anti-invariant combinations
$\Delta\mu_3$ and $\Delta\mu_4$ turn out to have vanishing
intersection form. Thus we find four mutually local dyons as it
should be for a $N\!=\!1$ vacuum.\\

Finally we note that in the last figure one sees a mirror pair of the
Dynkin diagram. This and the appearance of the Chebyshev polynomials
is obviously related to the resolution of simple singularities
\cite{arn}. It also allows for another beautiful interpretation.
Identifying the electric cycles around the branch cuts with weights
and the magnetic vanishing cycles with simple roots we recognize the
weight diagram of the fundamental representation of $D_4$!

%
%
\noindent {\bf Acknowledgements}

We would like to thank W. Lerche for drawing our attention
to this problem and for many helpful discussions.
This work has been supported by the BMWFK.

%
%

\end{document}